\documentclass[twocolumn,showpacs,amsmath,amssymb,prb,floatfix,showkeys]{revtex4}

\usepackage{graphicx}% Include figure files
\usepackage{dcolumn}% Align table columns on decimal point
\usepackage{bm}% bold math

\begin{document}

\title{Magnetic Anisotropy, Spin Pinning and Exchange Constants of (Ga,Mn)As films}

\author{Ying-Yuan Zhou, Yong-Jin Cho, Zhiguo Ge, Xinyu Liu, Malgorzata Dobrowolska, and Jacek K. Furdyna}
\affiliation{Department of Physics, University of Notre Dame, Notre Dame, IN 46556, USA}

\begin{abstract}
We present a detailed investigation of 
exchange-dominated nonpropagating spin-wave modes in a series of 100 nm 
Ga$_{1 - x}$Mn$_{x}$As films with Mn concentrations 
$x$ ranging from 0.02 to 0.08. The angular and Mn 
concentration dependences of spin wave resonance modes have been studied for 
both as-grown and annealed samples. Our results indicate that the magnetic 
anisotropy terms of Ga$_{1 - x}$Mn$_{x}$As depend on the 
Mn concentration $x$, but are also strongly 
affected by sample growth conditions; moreover, the magnetic anisotropy of 
Ga$_{1 - x}$Mn$_{x}$As films is found to be clearly linked 
to the Curie temperature. The spin wave resonance spectra consist of a 
series of well resolved standing spin-wave modes. The observed mode patterns 
are consistent with the Portis volume-inhomogeneity model, in which a 
spatially nonuniform anisotropy field acts on the Mn spins. The analysis of 
these exchange-dominated spin wave modes, including their angular 
dependences, allows us to establish the exchange stiffness constants for 
Ga$_{1 - x}$Mn$_{x}$As films.
\end{abstract}

%\begin{keywords}
%Exchange stiffness constant, ferromagnetic semiconductors, magnetic anisotropy, magnons, spin waves.
%\end{keywords}

\maketitle

\section{Introduction}
Incorporating Mn ions into III-V semiconductors makes it possible to achieve 
ferromagnetic order in semiconductor nanostructures \cite{Ohno:1998}. It 
is known that the spin waves in a magnetic system are 
determined by (and can be used to obtain) the exchange and magnetic 
anisotropy parameters \cite{nig:2001,Liu:2003,Rappoport:2004,
Goennenwein:2003a,Goennenwein:2003b}, and are thus particularly useful for a 
quantitative understanding of ferromagnetism in Ga$_{1 - x}$Mn$_{x}$As 
\cite{nig:2001}. In this paper we use ferromagnetic resonance 
(FMR), a powerful tool for investigating magnetic anisotropy and exchange 
constants in ferromagnetic materials \cite{Liu:2006}, to investigate a 
series of Ga$_{1 - x}$Mn$_{x}$As films with different Mn concentrations $x$. In 
all samples studied we observe a multi-mode spin wave resonance (SWR) 
spectrum. We will focus on the detailed description of such 
SWRs, and the relationship between these spin wave modes and 
the Mn concentration $x$ in Ga$_{1 - x}$Mn$_{x}$As.

\section{Sample Fabrication}

A series of Ga$_{1 - x}$Mn$_{x}$As films were grown by
molecular beam epitaxy (MBE) on semi-insulating GaAs (001) 
substrates. A 100 nm-thick GaAs buffer was first grown at the substrate 
temperature 600$^{\circ}$C to achieve an atomically flat surface. The substrates 
were then cooled to 250$^{\circ}$C for growth of a 2 nm low temperature GaAs 
buffer, followed by 100 nm Ga$_{1 - x}$Mn$_{x}$As layers with various Mn 
concentrations ($x$ = 0.02, 0.025, 0.04, 0.045, 0.055, 0.06, 0.07 and 0.08). 
The thickness of (Ga,Mn)As samples is specifically selected to reliably 
determine the effect of annealing on SWRs. All as-grown specimens exhibited 
ferromagnetic order, with Curie temperature 
$T_{C}$ ranging from 50 K to 80 K. Pieces of samples cleaved from each 
specimen were then annealed in N$_{2}$ gas for one hour at 280$^{\circ}$C in 
order to examine the effect of annealing on magnetic properties. 
As a result of annealing $T_{C}$ of the samples 
increased to a range from 55 K to 115 K. Note that $T_{C}$ follows a general 
rule for both as-grown and annealed samples: it increases with increasing 
concentration $x$. 

\section{Experimental Setup}
\label{sec:experimental}

In this study we have measured the angular dependence of FMR for each 
specimen (both as-grown and annealed) in three geometries 
\cite{Liu:2006}. FMR measurements were carried out at 9.46 GHz 
using a Bruker electron paramagnetic resonance spectrometer. The applied dc 
magnetic field \textbf{H} was in the horizontal plane, while the microwave 
magnetic field was acting vertically on the sample. The sample was placed in 
a suprasil tube inserted in a liquid helium continuous flow cryostat, which 
could achieve temperatures down to 4.0 K. 

The Ga$_{1 - x}$Mn$_{x}$As layers were cleaved into three square pieces with 
edges along the [110] and $[1\overline 1 0]$ directions. Each square piece 
was then placed in three different orientations. Geometry 1 is when the 
sample plane and the $[1\overline 1 0]$ edge are vertical, which allows 
measurement with dc magnetic field \textbf{H} oriented at any angle between 
\textbf{H}$\parallel$[001] (normal orientation) and \textbf{H}$\parallel 
$[110] (in plane orientation). Geometry 2 is when the sample plane 
and the [010] direction are vertical, which allows us to measure FMR with 
field orientations between \textbf{H}$\parallel$[001] and the in-plane 
orientation \textbf{H}$\parallel$[100]. Geometry 3 is when the sample 
plane is horizontal, allowing us to map out the FMR when \textbf{H} is 
confined to the layer plane.

\section{Results and Discussions}

\subsection{Magnetic Anisotropy} 
Multi-mode SWR spectrum was observed in all samples. Our analysis of the 
data was carried out as follows. We first obtained the magnetic anisotropies 
and $g$-factors by fitting the angular dependence of the 
strongest resonance line to a theoretical uniform FMR model using a 
nonlinear least squares method \cite{Liu:2005}. Our fitting results show 
that the magnitude of perpendicular uniaxial anisotropy $H_{2 \perp }$ has 
tends to increase with Mn concentration $x$, while the in-plane cubic 
anisotropy $H_{4\parallel}$ decreases with $x$. However, the relation 
between the magnetic anisotropy terms and $x$ is strongly 
influenced by the individual growth condition of each sample. Furthermore, 
we find that the term 4$\pi M_{eff}$ = 4$\pi M - H_{2 \perp  }$ \cite{Liu:2003} is linearly increasing with 
the Curie temperature $T_{C}$ in both as-grown and annealed specimens, and 
$H_{4\parallel}$ is monotonically decreasing with $T_{C}$, with the 
exception of annealed specimens with the highest values of $T_{C}$. The 
results are shown in Fig. 1. Since an earlier report 
\cite{MacDonald:2005} shows an empirical relationship $T_{C} \sim $ 
$p^{1 / 3}$, we immediately note that there exists a tight relationship 
between magnetic anisotropy and the hole concentration $p$. However, the data 
for as-grown and annealed samples follow different slopes as a function of 
$T_{C}$, suggesting that there exists a fundamental differences between 
as-grown and annealed samples regarding magnetic anisotropy. Nevertheless, 
the distinct characteristics which we observe may provide useful insights 
into the origin of magnetic anisotropy in Ga$_{1 - x}$Mn$_{x}$As films.

\begin{figure}
\centering
\includegraphics[width=0.72\linewidth]{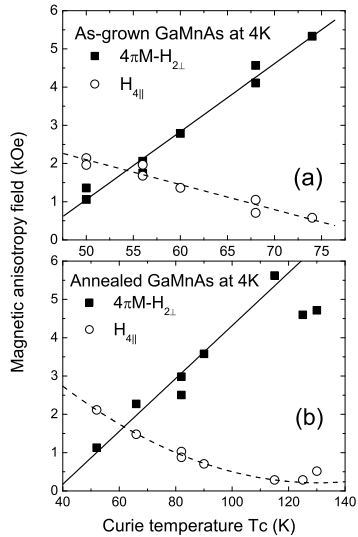}
\caption{The effective anisotropy term 4$\pi M-H_{2 \perp }$ and the in-plane 
cubic anisotropy $H_{4\parallel}$ versus the Curie temperature of (a) 
as-grown samples, and (b) annealed samples at T = 4 K. The solid and dashed 
lines are guides for the eye.}
\label{fig1}
\end{figure}

\subsection{SWR Spectra} 
It is the spin pinning at sample surfaces that induces spin-wave excitation 
in the FMR experiment. To investigate this feature in detail, we have chosen 
four specimens grown at similar conditions, with Mn concentration $x$ = 0.02, 
0.04, 0.06 and 0.07. Figure 2 shows the SWR spectra at T = 4 K for both 
as-grown and annealed samples when the dc magnetic field \textbf{H} is 
normal to the sample plane (\textbf{H}$\parallel $[001], i.e., $\theta 
_{H}$ = 0$^{\circ})$. In both cases the SWR spectra consist of several 
well-resolved standing spin wave modes separated by roughly equal field 
increments. Note that the SWR fields are increasing with the Mn 
concentration $x$, except in the case of one annealed sample with Mn 
concentration $x$ = 0.07. Importantly, the as-grown samples have a larger mode 
separation than the annealed samples, which indicates that the exchange 
stiffness constant $D$ is larger in the as-grown samples, as discussed later. 

\begin{figure}
\centering
\includegraphics[width=0.77\linewidth]{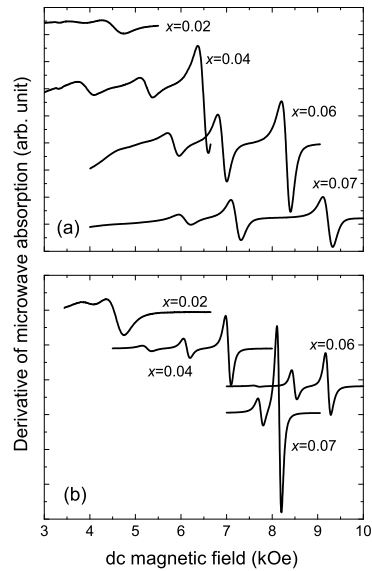}
\caption{The SWR spectra at 4 K when the external dc magnetic field is 
perpendicular to the film (\textbf{H}$\parallel $[001]) for (a) as-grown 
samples and (b) annealed samples.}
\label{fig2}
\end{figure}

As \textbf{H} is rotated away from the perpendicular orientation, the 
low-field spin wave modes gradually disappear, and eventually only one 
narrow resonance line remains at a critical angle $\theta _{c}$. We will 
refer to it as the uniform mode. For angle $\theta _{H} > \theta _{c}$ 
the spectrum generally consists of two wide resonance lines. We identify the 
mode lying at the higher field as an exchange-dominated surface spin wave 
mode \cite{Puszkarski:1979}, \cite{Liu:1}. The angular dependence of 
the SWRs reveals the nature of surface spin pinning and its dependence on 
the magnetization orientation. Using the Puszkarski surface inhomogeneity 
(SI) model \cite{Puszkarski:1979}, it is found that, as the magnetization 
rotates from the perpendicular to the in-plane orientation, the surface 
spins are evolving from a strong pinning condition to a weak pinning 
condition, with a turning point at the critical angle $\theta _{c}$ 
\cite{Liu:1}. Note that for \textbf{H}$\parallel$[001] the 
pinning is not only strong, but also nonlocalized, i.e., the magnetic spins appear to 
be affected by a spacially \textit{nonuniform} anisotropy field \cite{Portis:1963}, 
suggested by a linear mode separation at this orientation.

\subsection{Exchange Constants} 
We will now focus on the SWR spectrum obtained for \textbf{H}$\parallel 
$[001]. It is found that the positions of the SWR modes comply with a linear 
mode separation model for \textbf{H}$\parallel $[001] when a symmetrical parabolic magnetic 
anisotropy is assumed along the growth direction $z$ $(\vert z\vert \le $ 
$L$/2): $4\pi M^\ast (z) = 4\pi M^\ast (0)(1 - 4\varepsilon z^2/L)$\cite{Sasaki:2003}. 
Here $\varepsilon $ is the distortion 
parameter of the film, $L$ = 100 nm is the film thickness, and $4\pi M^\ast (0) 
= 4\pi M - H_{2 \perp } - H_{4 \perp } $. In this situation the position of 
the $n$-th SWR mode for \textbf{H}$\parallel $[001] is given by the Portis 
relation \cite{Portis:1963}:

\begin{equation}
\label{eq1}
H_n = H_0 - (n - \frac{1}{2})(4 / L)(4\pi M^\ast (0)\varepsilon 
\frac{D}{g\mu _B })^{1 / 2},
\end{equation}

\noindent
where \textit{$\mu $}$_{B}$ is the Bohr magneton, $n$ is an odd integer, 
and $H_0 = \omega/\gamma + 4\pi M^\ast (0)$ is the position of the 
theoretical uniform mode. Here $\omega $ is the angular frequency of the microwave 
field and $\gamma $ is the gyromagnetic ratio. The exchange stiffness constant $D$, 
which gives a measure of the strength of the exchange interaction, can be 
determined from the difference of the adjacent SWR modes by: 

\begin{equation}
\label{eq2}
\frac{D}{g\mu _B } = \frac{\Delta H_{n1,n2}^2 L^2}{16(n2 - n1)^2(4\pi M^\ast 
(0)\varepsilon )}.
\end{equation}

\noindent
Note that the value of 4\textit{$\pi $M}$^{\ast }$(0)\textit{$\varepsilon $}, the depth of parabolic potential 
well, can be roughly evaluated from the field separation between the highest 
and the lowest SWR modes observed for \textbf{H}$\parallel $[001] \cite{Liu:1}.
In the special case where only two modes are observed (see Fig. 2, $x$ = 0.07), 
it is interesting that $\Delta H_{1,3} \approx 4\pi M^\ast (0)\varepsilon$, so that $D$ 
can be satisfactorily estimated simply from the separation of the two modes.

Using Eq. (\ref{eq2}), we can obtain the value of $D$/\textit{g$\mu $}$_{B}$ for all samples. 
The data for four optimally-grown samples are plotted as function of Mn 
concentration $x$ in Fig. 3. The figure clearly shows that $D$ is closely related to $x$. In 
particular, for as-grown samples $D$ decreases very quickly as $x$ increases. On 
the other hand, although the data for annealed samples suggest that $D$ also 
decreases as $x$ increases, the change is much smaller in this case. The most 
important characteristic shown in Fig. 3 is that $D$ decreases after annealing, 
especially for low Mn concentration samples. This experimental result is 
quite surprising, because $T_{C}$ -- which usually go 
hand-in-hand with the strength of the exchange interaction -- are higher in 
the annealed samples than in the as-grown films. One possible explanation 
is that the average distance between substitutional Mn ions (which links $T_C$ 
to $D$) is changed by the annealing process sufficiently to cause such a result. 
Indeed, significant changes in the lattice parameter of (Ga,Mn)As have been 
observed experimentally. However, the exact mechanism of such behavior 
is not understood and requires further study.

\begin{figure}
\centering
\includegraphics[width=0.8\linewidth]{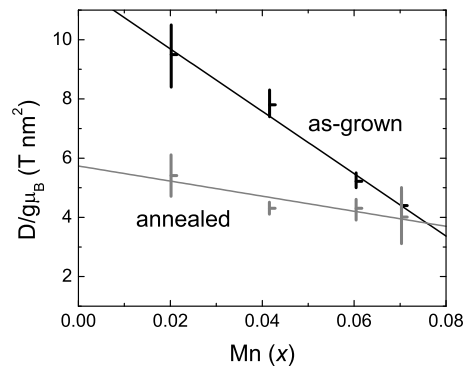}
\caption{The exchange stiffness constant $D$ as a function of Mn concentration 
$x$ for as-grown and annealed GaMnAs films. The lines are linear fits.}
\label{fig3}
\end{figure}

\section{Summary and Conclusions}
In summary, we carried out a detailed experimental study of SWRs in thin 
Ga$_{1 - x}$Mn$_{x}$As films with a wide range of Mn concentrations $x$. The 
analysis of the data allowed us to establish the following important 
magnetic properties of these films: magnetic anisotropy, surface spin 
pinning, and the exchange stiffness constant. The magnetic parameters 
obtained from this analysis were studied as function of $x$ 
and/or $T_{C}$ of the specimens. We observe a significant 
change of magnetic properties between as-grown and annealed samples. The 
results clearly show that the study of SWRs provides valuable information 
about the strong correlation between structural and magnetic properties of 
these materials.

\section*{Acknowledgment}
This work was supported by the NSF Grant DMR06-03752.

\end{document}